\documentclass[conference]{IEEEtran}
\IEEEoverridecommandlockouts

\usepackage{lipsum}
\usepackage{cite}
\usepackage{amsmath,amssymb,amsfonts}
\usepackage{algorithmic}
\usepackage{graphicx}
\usepackage{textcomp}
\usepackage{xcolor}
\usepackage{epsfig}
\usepackage{epstopdf}
\usepackage{balance}
\usepackage{caption}
\usepackage{subcaption}
\usepackage[printonlyused]{acronym}
\usepackage[T1]{fontenc}
\usepackage{booktabs}
\usepackage{ragged2e}
\usepackage{array}
\usepackage{tabularx}

\acrodef{qos}[QoS]{quality--of--service}
\acrodef{cpe}[CPE]{customer premises equipment}
\acrodef{poe}[PoE]{power--over--ethernet}
\acrodef{kpi}[KPI]{key performance indicator}
\acrodef{rsrp}[RSRP]{Reference signal received power}
\newcommand{\FGR}[1]{Fig.~\ref{#1}}

\newcommand{\TAB}[1]{Table~\ref{#1}}
    
\begin{document}

\title{A Comparative and Measurement-Based Study on Real-Time Network KPI Extraction Methods for 5G and Beyond Applications\\
}

\author{
\IEEEauthorblockN{
Batuhan KAPLAN,
Samed KEŞİR,
Ahmet Faruk COŞKUN
}

\IEEEauthorblockA{6GEN Laboratory, Next-Generation R\&D, Network Technologies, Turkcell, İstanbul, Türkiye}

Emails: \{batuhan.kaplan, samed.kesir, coskun.ahmet\}@turkcell.com.tr

}

\maketitle

\begin{abstract}
Key performance indicators (KPIs), which can be extracted from the standardized interfaces of network equipment defined by current standards, constitute a primary data source that can be leveraged in the development of non-standardized new equipment, architectures, and computational tools. In next-generation technologies, the demand for data has evolved beyond the conventional log generation or export capabilities provided by existing licensed network monitoring tools. There is now a growing need to collect such data at specific time intervals and with defined granularities. At this stage, the development of real-time KPI extraction methods and enabling their exchange between both standardized/commercialized and non-standardized components or tools has become increasingly critical. This study presents a comprehensive evaluation of three distinct KPI extraction methodologies applied to two commercially available devices. The analysis aims to uncover the strengths, weaknesses, and overall efficacy of these approaches under varying conditions, and highlights the critical insights into the practical capabilities and limitations. 
The findings serve as a foundational guide for the seamless integration and robust testing of novel technologies and approaches within commercial telecommunication networks. This work aspires to bridge the gap between technological innovation and real-world applicability, fostering enhanced decision-making in network deployment and optimization.
\end{abstract}


\section{Introduction}

In the rapidly evolving world of telecommunications, the emergence of 5G networks marks a transformative milestone, redefining connectivity and operational efficiency. These networks promise improved security, coverage, reliability, and connected device density, and also support the adoption of networks by new application areas and use cases \cite{1}. In the context of expanding communication networks across vertical sectors and application domains, flagship projects focusing on preliminary demonstrations, experimentations and testbed applications hold significant importance. Particularly in application areas with a strong focus on innovation, demonstration studies often face challenges due to the integration of recently proposed, yet-to-be-standardized technological concepts (e.g., physical layer prototypes, artificial intelligence/machine learning (AI/ML) approaches) into the existing commercial networks. Here, the primary issue arises from the inability to establish data flow—or to do so in a manner that meets the required data access frequency and/or resolution—between commercially available products, whose interfaces and communication protocols are predefined, and innovative technological components that have not yet reached commercial maturity. For applications requiring real-time processing of instantaneous data obtained from physical interfaces (e.g., Ethernet, COM port) at specific nodes in the network, existing access methods alone are insufficient. At this stage, hybrid approaches leveraging the advantages of network monitoring tools, protocols, and open-source software libraries come to the forefront and can be utilized to obtain \acp{kpi} experienced in network equipment. The primary and commonly referred network \acp{kpi} include received signal strength indicator (RSSI), \ac{rsrp}, reference signal received quality (RSRQ), and signal-to-interference plus noise ratio (SINR). These \acp{kpi} are crucial for assessing network quality and performance, influencing decisions related to handovers, link adaptation, and overall network management \cite{2, 3, 4, 5, 6}, and they are also integral to radio resource management in 5G networks, providing information on signal strength and quality \cite{3, 5, 7, 8}.

\begin{figure*}[!h]
    \centering
    \begin{subfigure}[b]{0.65\textwidth}
        \centering
        \includegraphics[width=\linewidth]{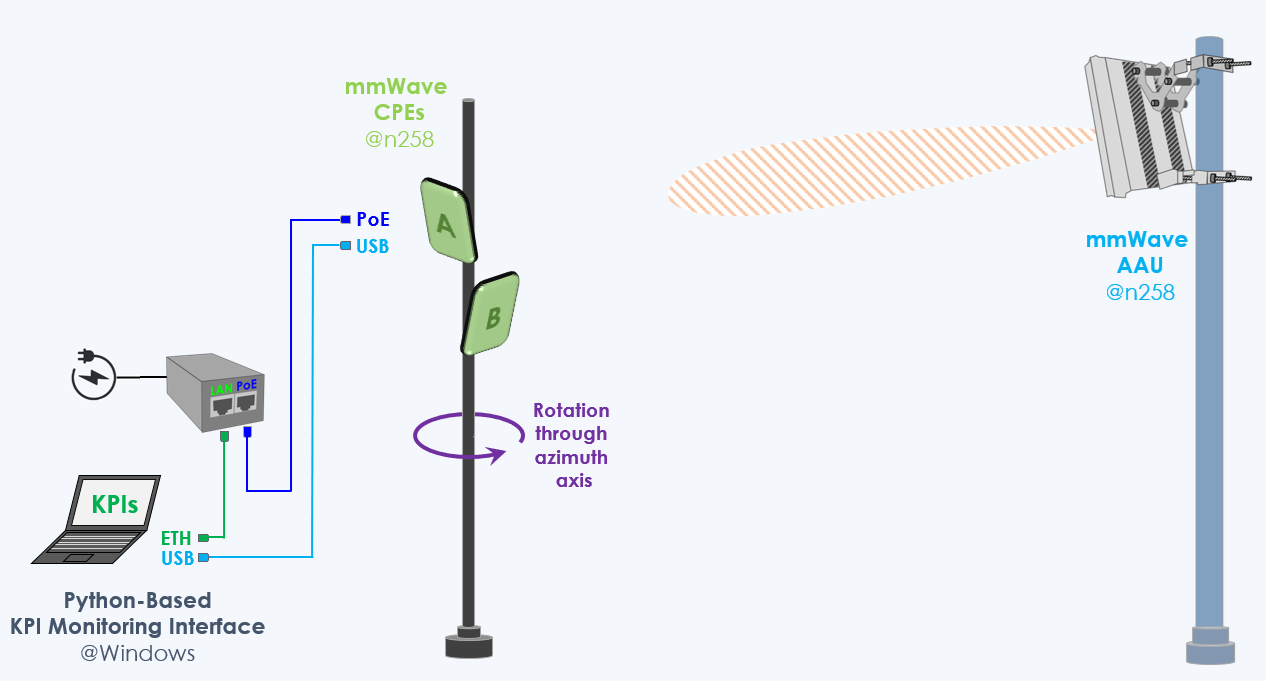}
        \caption{Schematic representation}
        \label{fig:Footage}
    \end{subfigure}
    \hfill
    \begin{subfigure}[b]{0.26\textwidth}
        \centering
        \includegraphics[width=\linewidth]{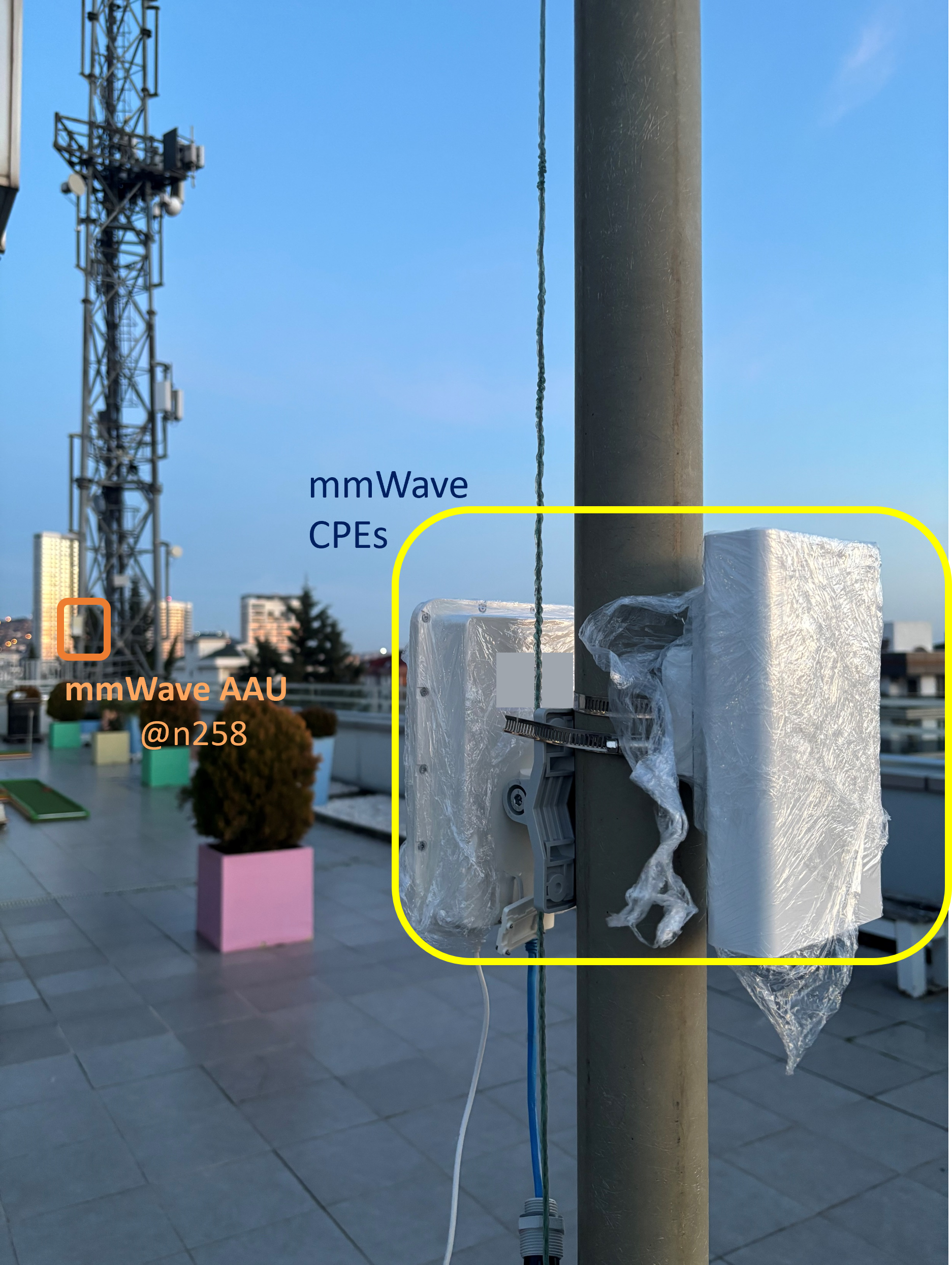}
        \caption{Footage}
        \label{fig:SchematicView}
    \end{subfigure}
    \caption{Description of the measurement setup}
    \label{fig:MeasurementSetup}
\end{figure*}

The extraction and analysis of these \acp{kpi} are pivotal for optimizing 5G network performance. Advanced techniques, such as data-driven optimization and ML, are increasingly employed to enhance the accuracy and efficiency of \ac{kpi} measurement and interpretation \cite{9, 10}. These methods enable network operators to automate the optimization of network parameters, thereby improving coverage, quality, and user experience \cite{5, 9}. Moreover, integrating AI frameworks for \ac{kpi} monitoring and diagnosis further enhances the ability to detect and address network issues proactively \cite{10}. Besides, the practical method involves using platforms like the SIM8200EA-M2 to connect with mobile terminals, enabling the collection and analysis of real-time data utilizing the AT commands \cite{2}. This approach not only facilitates the validation of 5G network parameters but also aids in understanding the intricate dynamics of cellular communication systems. Data extraction from 5G networks is a critical yet challenging task, often requiring labor-intensive and costly measurement campaigns. \cite{11} addresses this issue by proposing 5GNN, a graph neural network (GNN)-based framework for extrapolating 5G signal metrics from sparse measurements. The authors conducted extensive field campaigns using Accuver XCAL to collect high-precision geolocated datasets of 5G signals. Among the studies in the literature focused on data extraction from nodes in commercial 5G networks, the two most comprehensive works are presented in \cite{2} and \cite{11}. The authors of \cite{1} perform data extraction in real-time using AT commands. In contrast, \cite{11} employs the commercial network monitoring software Accuver XCAL, which focuses on recording observed KPIs with timestamping during communication rather than operating a real-time process. For this purpose, the software's logging feature is utilized; however, real-time access to specific KPI datasets is not achieved.

In this study, a comprehensive analysis of three different data access approaches is conducted to reveal the capabilities and limitations of the methods under consideration. The ultimate aim is to provide a detailed guide for the integration and testing of innovative technologies and approaches within commercial telecommunication networks.

\section{Experimental Setup on Commercial 5G Wireless Deployment}

A schematic representation and photographic visual of the 5G communication scenario, which has been set up on the terrace of Turkcell Kartal Plaza using commercial telecommunication products to implement the 5G network KPI extraction approaches to be examined within the scope of this study, are provided in \FGR{fig:MeasurementSetup}. The following subsections respectively provide key information for the measurement methodology, network specifications and equipment, and the utilized KPI extraction methods to give the readers the background knowledge about the entire network KPI extraction process. 

\begin{table*}
\vspace{0.8382mm}
\centering
\caption{Comparison of Query Results For Methods}
\vspace{-0.8382mm}
\label{tab:responseofmethods}
\setlength{\arrayrulewidth}{0.25mm}
\def\arraystretch{1.25}
\resizebox{\textwidth}{!}{%
\begin{tabular}{c|cc|cc|c}
\hline\hline
\textbf{Parameters} & \textbf{GM Web Interface} & \textbf{Meig Web Interface} & \textbf{AT\textasciicircum{}DEBUG Query} & \textbf{\begin{tabular}[c]{@{}c@{}}AT+SGCELLINFOEX\\ Query\end{tabular}} & \begin{tabular}[c]{@{}c@{}}\textbf{Accuver XCAL}\\ x80A3 Command\end{tabular} \\ \hline\hline
\textbf{RAT} & 5G & 5G & NR5G\_SA & 5G & 5GNR \\ \hline
\textbf{MCC, MNC} & 286, 01 &  & 286, 01 & 286, 01 &  \\ \hline
\textbf{NR Cell ID} & 16400395 &  & 16400395 & 16400398 &  \\ \hline
\textbf{NR TAC} &  &  & 1000 & 1000 &  \\ \hline
\textbf{Physical Cell ID} & 2 & 2 &   & 2 & 2 \\ \hline
\textbf{Band} & n258 & n258 & n258 & 258 & 258 \\ \hline
\textbf{Bandwidth} & 200.0 MHz &  & 200.0 MHz & 200 MHz &  \\ \hline
\textbf{Sub-Carrier Spacing} &   &  &   & 120 & 120 kHz \\ \hline
\textbf{Frequency Range Type} &  &  &   & 2 &  \\ \hline
\textbf{DL/UL Channel} & 2058427 &  & 2058427 & 2058427 &  2058427 \\ \hline
\textbf{RSSI} &   &  & 3 (-84.3 dBm, -78.6 dBm,  ,  ) &  &  \\ \hline
\textbf{RSRP} & -78.1 dBm & -80 dBm & -79 dBm & -80 dBm & -78.02 dBm \\ \hline
\textbf{RSRQ} & -11.6 dB & -11 dB & -11 dB & -11 dB & -11.17 dB \\ \hline
\textbf{SNR} & 12 dB &  &  &  &  \\ \hline
\textbf{SINR} &  & 12.0 dB & 14.0 dB & 14.5 dB & 14.21 dB \\ \hline
\textbf{Duplex Mode} &  &  &   & TDD NR5G & TDD \\ \hline\hline
\end{tabular}%
}
\end{table*}

\subsection{Methodology}
The active antenna unit (AAU) is mounted on a huge antenna pole and has a height of $3.5$ m with respect to the terrace level. As seen from the test setup visuals in Fig. \ref{fig:MeasurementSetup}, two different commercial \ac{cpe} products from two vendors have been utilized for the KPI extraction process. The CPEs with a distance of $30$ m to AAU are mounted on a tripod at a height of $1.7$ m from the terrace level where the tripod can rotate in the azimuth direction.

The CPE products have been utilized as end-user equipment (UE) that will constitute the main access nodes. The extraction procedures that will be introduced in Section II-C, are based on accessing the CPE devices via their common data\&control interfaces (e.g., Ethernet, USB, etc.) and access protocols (e.g., HTTP, SSH, Telnet, etc.) and processing the response of the devices that include the measured network KPIs. Here, the execution of remote access, control, data parsing, and serialization processes are performed by the Python libraries{\footnote{The repository of this study including the Python libraries for KPI extraction methods is available online: https://github.com/kesirsamed/KPI-extraction-from-Customer-Premises-Equipments.}} developed by the authors in addition to some open-source ones. As seen from Fig. \ref{fig:MeasurementSetup}-(a), both CPEs provide their DC power from the \ac{poe} injector and can be accessed by the UE laptop via the common interfaces.

In order to incorporate a device-specific perspective into the comparison of the proposed KPI extraction methods, these three methods were evaluated for both CPE devices. The performances of the device responses to these three methods are compared in terms of KPI resolution and data update frequency. Since the evaluations are conducted within the framework of a highly specific scenario and some message sets/access techniques, to avoid any commercial bias in the comparison of the CPE brands, the devices are referred to as CPE-A and CPE-B in the results section, despite the technical details being explicitly provided in subsequent sections.

\subsection{Network and Equipment Specifications}
This subsection briefly exhibits the basic specifications of the utilized 5G infrastructure and its components.

\subsubsection{AAU and Base Station}
The AAU operates in the n258 mmWave band (27.1–27.5 GHz) and supports 5G New Radio (NR) technology. Despite owning up to 64 beams for advanced beamforming and spatial multiplexing, for this study, the AAU was configured to operate in a single-beam mode to mitigate possible beam switching due to channel fluctuations. The radio access components have been served by the mobile private network core solutions of Huawei Technologies Co. Ltd. which has a 5G standalone (SA) architecture. 

\subsubsection{CPE Devices}
The device with the name MEIGCEE SRT853L Outdoor \ac{cpe} \cite{12} has been powered by the Qualcomm SDX65 chipset \cite{13}, which delivers advanced 5G connectivity with support for both the mmWave and Sub-6 GHz spectrum bands. It achieves download speeds of up to $10$ Gbps and upload speeds up to $3.38$ Gbps on mmWave. Equipped with internal QTM547 antenna modules, it ensures reliable performance and supports SA and NSA network modes. 

Another \ac{cpe} device is the General Mobile OD-513 \cite{14}, powered by Qualcomm's Snapdragon X55 chipset \cite{13}, offers advanced global 5G connectivity over both mmWave and sub-6 GHz bands. It achieves download speeds up to $7$ Gbps and upload speeds up to $3$ Gbps while providing multi-mode support across 5G NR, LTE and legacy 3G and 2G networks.

\subsection{KPI Extraction Methods}
In 5G network \ac{kpi} measurements, there are two primary approaches: active and passive \cite{15}. Active methods involve generating traffic or simulating user behavior to evaluate network performance without requiring root access to the mobile device. These methods are user-friendly and can be implemented on standard devices. Passive methods rely on monitoring the actual traffic and system performance at a deeper level. These require root access to the mobile device, \ac{cpe} in our case, providing a more detailed insight into the network metrics but often at the cost of increased complexity and reduced device compatibility. In this study, three passive approaches are adopted, and their corresponding tools are utilized throughout the measurements. The details of each method discussed are provided in sequential subsections, while the list of network KPIs that could be captured from the devices introduced in Section II-B-2 is presented in \TAB{tab:responseofmethods}.

\subsubsection{Web GUI-Based KPI Extraction}
The first and simplest method for KPI extraction is the use of internet protocol (IP)--based web interfaces, which allow the network devices to be monitored or configured externally. With this method, various network KPIs exhibited in tabular form or as plain text within specific tabs/pages and sections of the graphical user interface (GUI) are accessible for almost all commercial devices. In this study, the GUIs encountered when accessing two examined CPE devices via Ethernet port using their IP addresses are presented in \FGR{fig:webfootage}.
To transform this approach into an automatic KPI monitoring method, it is necessary to navigate to the page/tab area where the mentioned KPI values are displayed via mouse clicks and continuously read the fields that contain the KPI values in the HTML content while performing the required parsing and processing steps. As the prescribed procedure exceeds human processing speeds, the Selenium tool has been employed to automate the process. 

This method begins with setting up Selenium WebDriver to imitate user interactions with the browser. First, the WebDriver is initialized and pointed to the device's web interface by inputting the IP address. Next, authentication credentials are provided, if required, to log into the interface. Once authenticated, the areas of the interface that hold KPI data are identified using one of the HTML elements, XPath selectors. Selenium WebDriver then interacts with these parts to extract relevant data, such as signal strength, throughput, and latency metrics. With Python's Selenium library, it is possible to easily retrieve all KPIs published on the device's Web GUI in real time with a very limited Python scripting cost. The KPIs that can be read from the web interfaces of both CPEs are listed in Table \ref{tab:responseofmethods}. As shown, RSRP, RSRQ, and SINR, which are essential for evaluating the performance of communication systems, are commonly provided. At this point, it has been observed that the KPIs obtainable from the web GUI are highly dependent on the design preferences of the vendor company.

\begin{figure}
    \centering
    \begin{subfigure}[b]{0.45\textwidth}
        \centering
        \includegraphics[width=\linewidth]{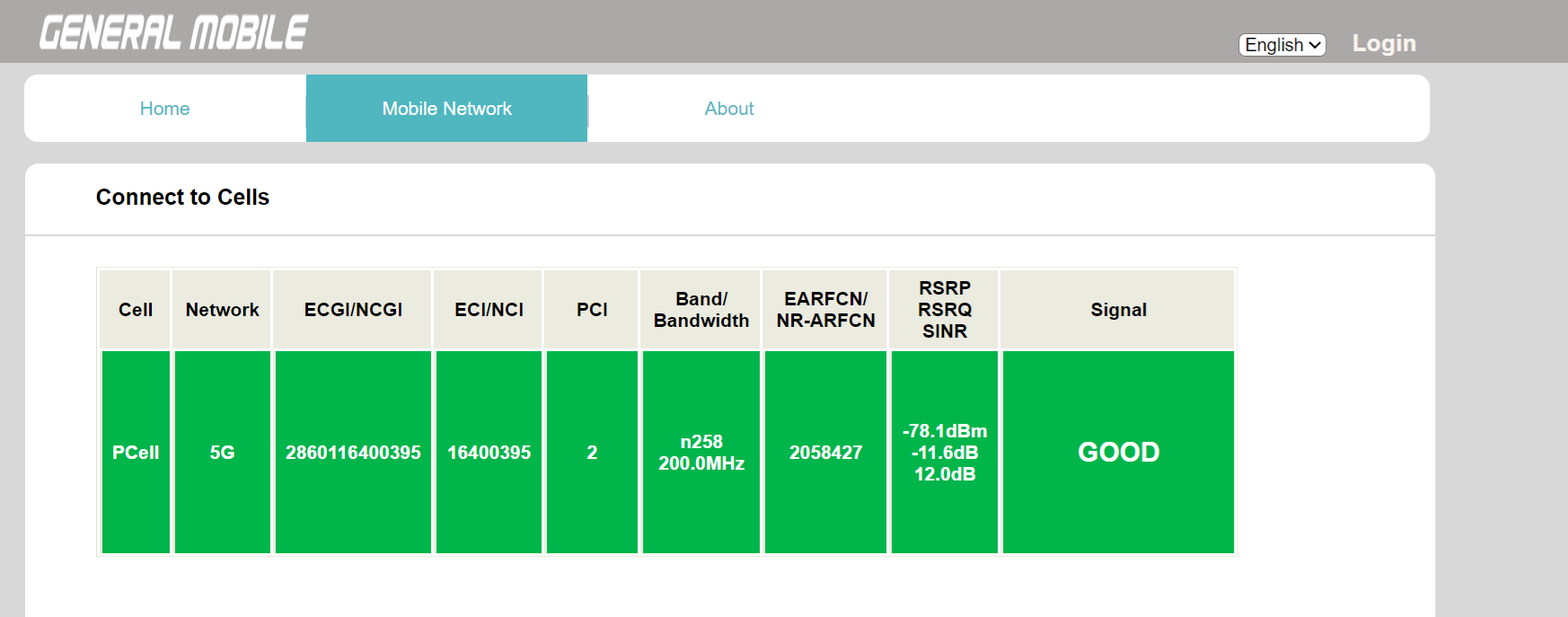}
        \caption{GM CPE web interface}
        \label{fig:gmodu-web-interface}
    \end{subfigure}
    \begin{subfigure}[b]{0.45\textwidth}
        \centering
        \includegraphics[width=\linewidth]{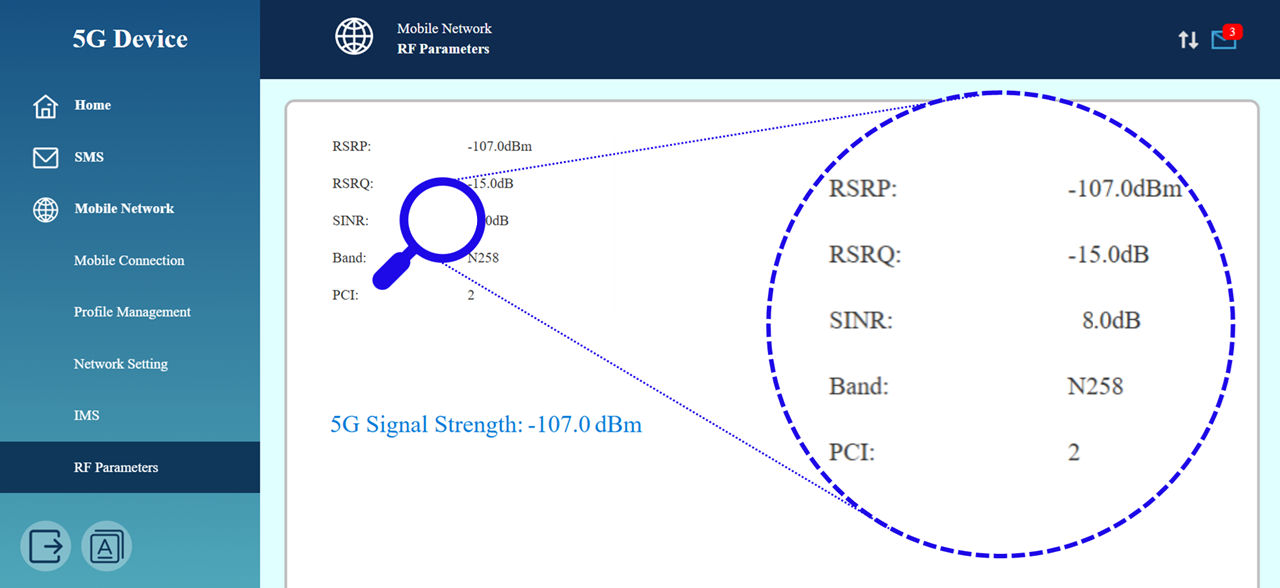}
        \caption{MeiG CPE web interface}
        \label{fig:meig-web-interface}
    \end{subfigure}
    \caption{Dashboards of the CPE devices for cellular information.}
    \label{fig:webfootage}
\end{figure}

\subsubsection{AT Commands-Based KPI Extraction}

A more widely used and license-free option for accessing network devices is based on utilizing AT queries. The AT commands sent by a serial connection over the COM port of the UE computer to the device, constitute a low-level interface for extracting key network metrics directly from the hardware. This method is highly efficient for detailed and custom performance analysis, especially in scenarios requiring specific diagnostic data. The entire AT command pool includes many different command strings that aim to interrogate different sets of network KPIs. Moreover, AT commands can vary from device to device due to differences in the included chipsets. Both CPE devices respond to only one of the AT commands used to query KPI metrics, "AT\textasciicircum DEBUG?" or "AT+SGCELLINFOEX?", while the other command does not generate a response. Apart from this distinction, the Android Debugging Bridge (ADB) shell is required to execute AT commands for one of the CPEs, while they can be executed directly via the terminal for the other. 
Both queries provide KPIs such as Cell ID, bandwidth, and RSRP. Thus, after establishing a connection between the UE laptop (for monitoring and evaluation purposes) and the CPE devices via serial communication (or IP-based communication over the Ethernet port), it becomes possible to send the aforementioned AT commands to the devices using Python's pyserial library functions. The parsed responses from the devices can then be processed to extract each KPI value. Table \ref{tab:responseofmethods} exhibits the KPIs that can be extracted after processing the queries' responses. As seen, an extensive list of network information and KPIs is provided by AT commands where the responses might differ from vendor to vendor.

\subsubsection{KPI Extraction w/ Accuver XCAL and TM}

Accuver XCAL \cite{16} stands out as a licensed software solution tailored for cellular network performance evaluation. It offers comprehensive functionalities for both active and passive measurement approaches, enabling detailed monitoring and analysis of various \acp{kpi} such as latency, throughput, and signal quality. With its specialized features and user-friendly interface, Accuver XCAL has become a widely adopted tool in the telecommunications industry, catering to both commercial and research needs in network performance evaluation. 

\begin{figure}
    \centering
    \begin{subfigure}[b]{0.99\linewidth}
        \centering
        \includegraphics[width=0.96\linewidth]{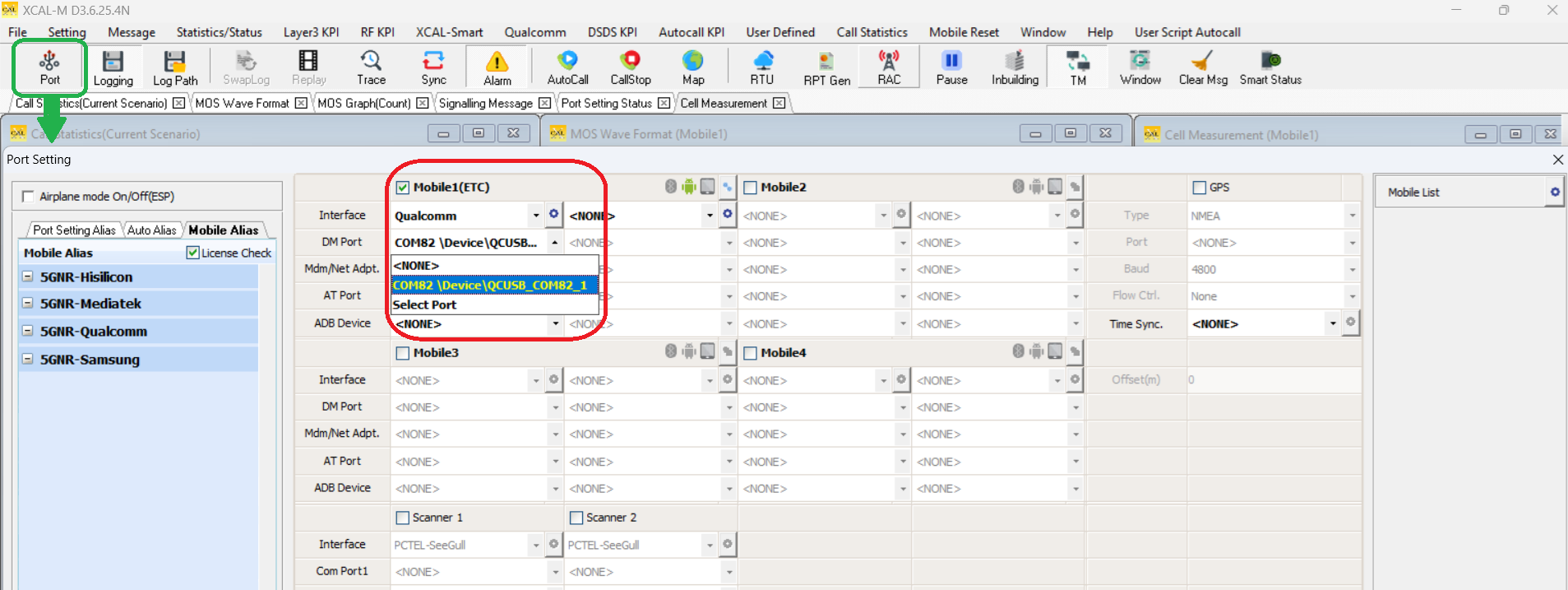}
        \caption{Port settings for device connection in Accuver}
        \label{fig:settingsaccuver}
    \end{subfigure}
    
    \begin{subfigure}[b]{0.99\linewidth}
        \centering
        \includegraphics[width=0.96\linewidth]{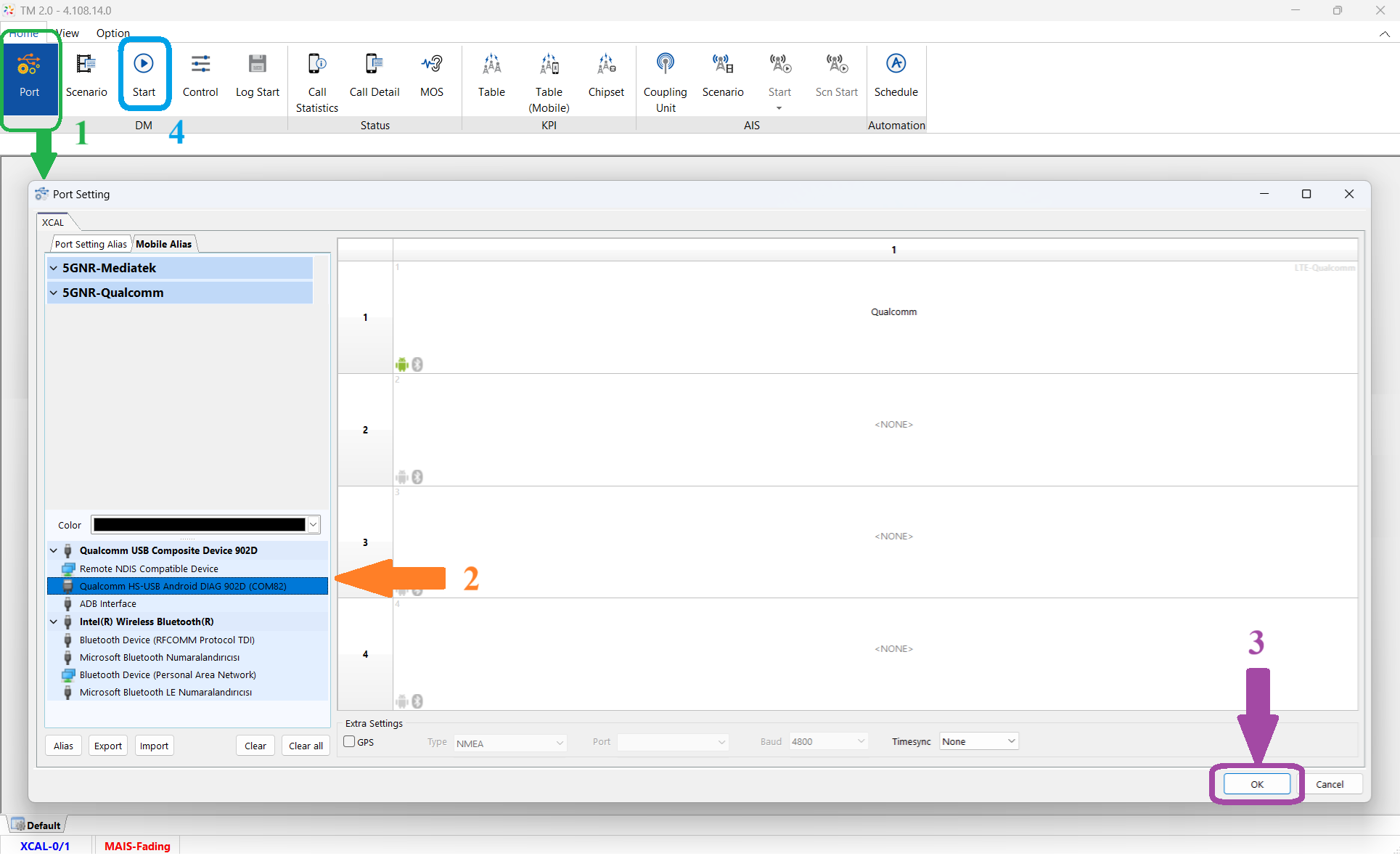}
        \caption{Autocall procedure defined between TM and Accuver}
        \label{fig:testManager}
    \end{subfigure}
    
    \begin{subfigure}[b]{0.99\linewidth}
        \centering
        \includegraphics[width=0.96\linewidth]{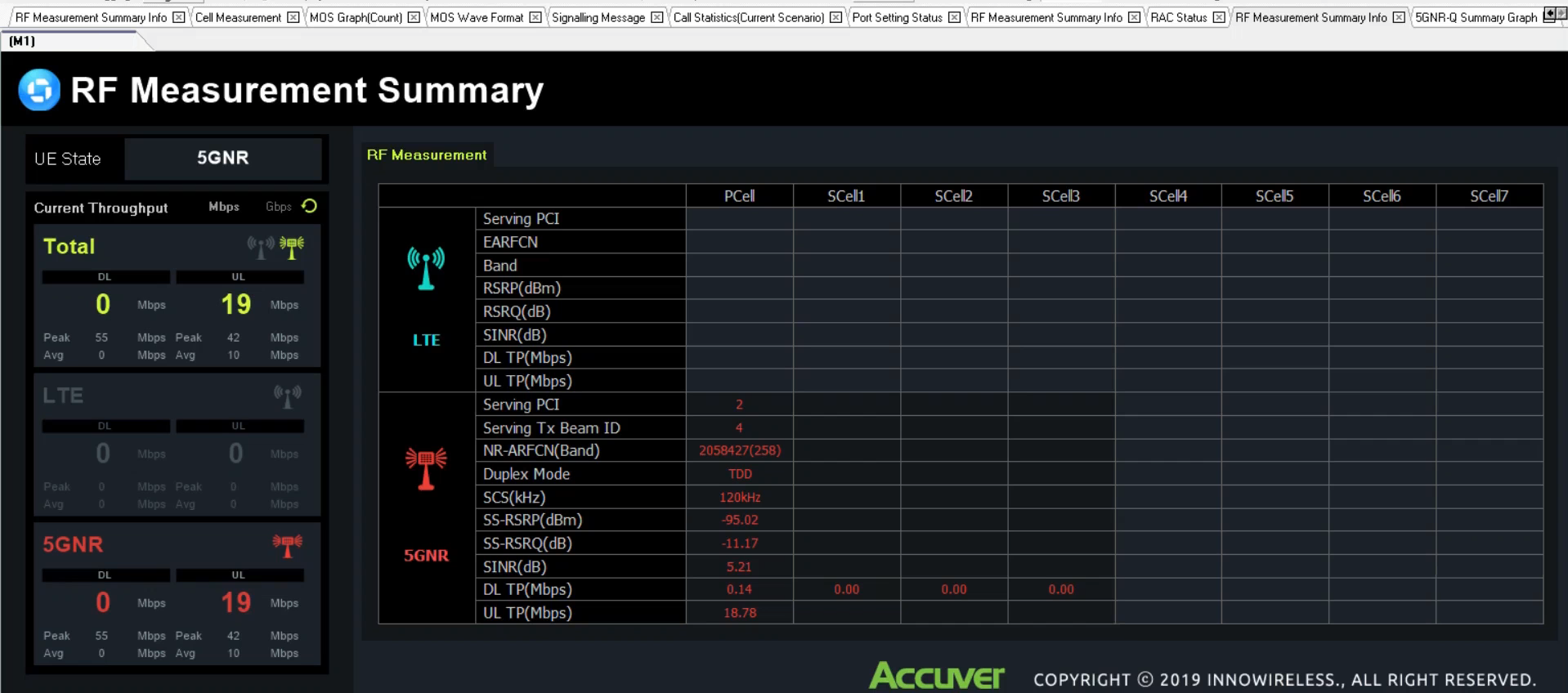}
        \caption{Accuver RF Measurement Summary (for a successful connection)}
        \label{fig:accuversummary}
    \end{subfigure}
    \caption{The preliminary settings for Accuver and TM}
    \label{fig:accuverscreenshots}
\end{figure}

Accuver solutions, which enable the performance monitoring of communication devices with various chipsets, include a plugin called Test Manager (TM) that facilitates the control and automation of systematic and simultaneous testing with multiple devices. Through this plugin, KPI extraction and monitoring processes can be customized to specific application requirements. Particularly in research-oriented projects, software development in various programming languages can be performed to control the TM tool that automates Accuver. For this study, the code development process has been carried out in Python using the socket connection, and a tailored remote access control (RAC) structure has been implemented. Before executing Python scripts with RAC authority, certain configurations need to be made through the Accuver and TM interfaces, such as selecting predefined ports and chipset information. Initially, the settings of the Accuver XCAL are configured for proper network connection as shown in \FGR{fig:settingsaccuver}. Also, \FGR{fig:testManager} shows the TM interface while the compatible device to start the autocall is selected. When autocall is started, Accuver XCAL starts to retrieve KPIs as shown in \FGR{fig:accuversummary}. Working with TM, Accuver XCAL can publish the data over the port in the local network. Here, among the RAC command list defined for Accuver, \texttt{0x80A3} command can be used to request the L3 KPI parameter report. Thanks to Python's socket library, in response to the \texttt{0x80A3} request sent according to the TCP protocol, the KPI content retrieved from Accuver is packaged by TM. As a result, the hexadecimal formatted response received using the socket library is processed using the built-in functions for ASCII to hex conversions, and the KPI parameters and values, provided in the rightmost column of Table \ref{tab:responseofmethods}, are obtained. It should be noted that the described KPI reading mechanism is common for both CPE devices. Additionally, this process is expected to run successfully for all devices with chipsets compatible with Accuver XCAL software. In Table \ref{tab:responseofmethods}, the common KPIs retrieved from Accuver are listed. Additionally, in response to the \texttt{0x80A3} request, the L3 KPI parameter report provides access to various 5GNR parameters such as frequency, synchronization signal block index, modulation and coding scheme, bit-error rate, etc.

\section{Measurement Results}

This section exhibits and compares the performance of the KPI extraction methods in terms of both KPI resolution and data update frequency for the 5G mmWave communication scenario described above. For the analysis, a 3-meter tripod capable of rotation along the azimuth axis, as depicted in the schematic illustration in \FGR{fig:MeasurementSetup}, is rotated at a specific speed for $30$ seconds. This enables variations in the KPIs to be observed based on the horizontal radiation pattern of the CPE devices. During this process, the communication KPIs experienced by both CPE devices are recorded and compared for all three methods, which has allowed to interpret how each method captures the changes in the KPIs in terms of both value resolution and refresh rate. For KPI acquisition, the UE laptop sends requests every $0.25$ seconds to retrieve network KPIs via each method so that the refresh rate of the methods can be observed. Here, just to provide a simple and direct comparison among the methods, only the RSRP parameter is used.

\FGR{fig:CPEA-rsrp-results} shows the variations of the RSRP for each method as \ac{cpe}-A rotates. The results demonstrate significant differences in the performance of these methods regarding time resolution, RSRP level resolution, and refresh rate in case of variations in signal conditions, highlighting their strengths and limitations for different devices. By examining the RSRP records of all methods, it can be seen that the Accuver-based method consistently outperforms the others, delivering the most stable and accurate results. Despite the higher signal variability observed during the rotation of the \ac{cpe}, the Accuver-based method effectively captures fine-grained signal changes while maintaining stability. Besides, the response of the AT command is seen to capture the effect of the rotation precisely despite having a reduced value resolution due to integer-conversion. The web interface is clearly seen to provide a coarse representation of the effect of rotation, with responses that exhibit a noticeable delay in tracking the changes. Furthermore, the data refresh rate differs between methods, although the request time has been kept the same for each. The AT method exhibits faster response times compared to the Accuver method, since response times are $0.25$ seconds and $1$ seconds for AT command and Accuver, respectively. On the other hand, although web interface can provide KPI parameters after every request sent, the returned KPIs lack to represent the physical RF conditions. Hence, this vulnerability should be assessed carefully for the applications to fulfill the data freshness and resolution requirements.

\begin{figure}
    \centering
    \begin{subfigure}[b]{0.45\textwidth}
        \centering
        \includegraphics[width=\linewidth]{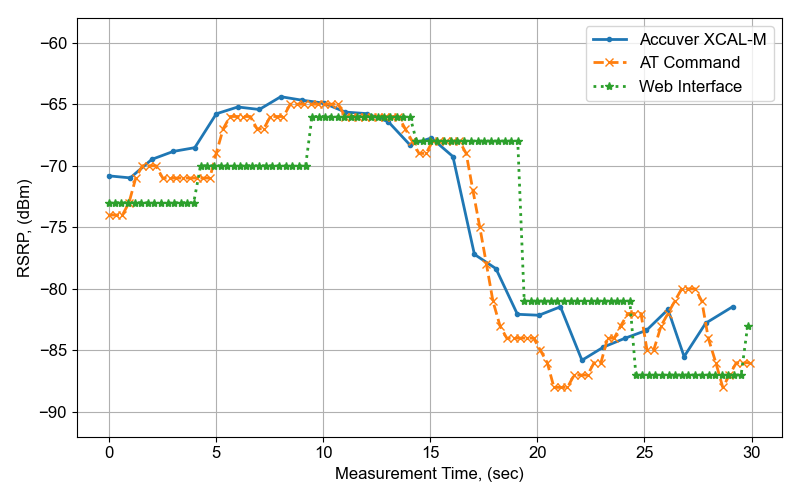}
        \caption{CPE-A}
        \label{fig:CPEA-rsrp-results}
    \end{subfigure}
    \begin{subfigure}[b]{0.45\textwidth}
        \centering
        \includegraphics[width=\linewidth]{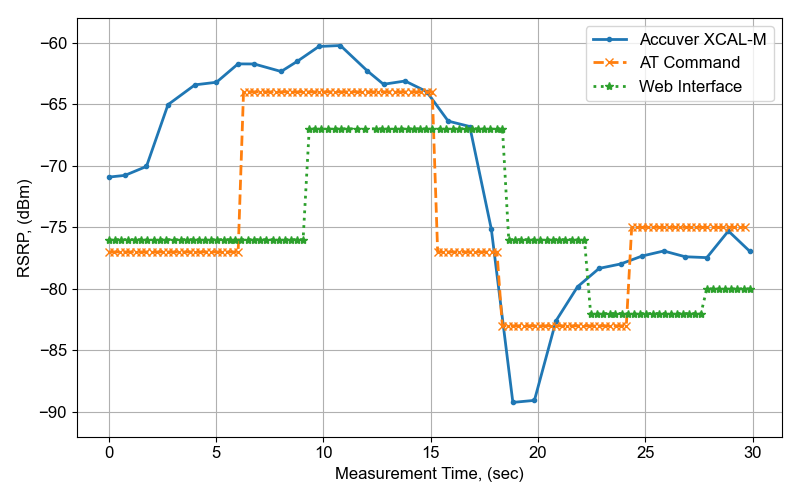}
        \caption{CPE-B}
        \label{fig:meig-rsrp-results}
    \end{subfigure}
    \caption{RSRP variations due to the rotation of CPEs}
    \label{fig:RSRPunderRotation}
\end{figure}

The analyses conducted for CPE-A were repeated for CPE-B, and as a result, the RSRP values obtained using three different KPI extraction methods were visualized in \FGR{fig:meig-rsrp-results}. Here, the XCAL-based method is seen to achieve superior accuracy and consistency due to its rapid update capability during the rotation, and providing floating point values. Whereas, the responses of the AT command and web interface for the RSRP are obtained in integer form. Additionally, although AT and web interface methods offer faster responses in case of consecutive requests, they both lack to capture the dynamic signal variations in different vulnerability levels. Therefore, AT and web interface methods cannot be suitable for time-sensitive applications in the case of CPE-B. 

\section{Conclusion}
Within the extensive study, the \ac{kpi} extraction methods, which are relying on web interface, AT commands, and Accuver XCAL software have been automated with the help of relevant Python libraries. Devices from two different vendors have been selected as the \acp{cpe} to serve as test points for evaluating the methods. This selection allowed for both the evaluation and elimination of device dependency in the methods, as well as the analysis of variations in two key metrics—KPI value resolution and refresh rate—during the KPI extraction process across different devices.
With the help of a simple measurement scenario with an azimuth rotation in the CPE pole, a time-varying channel condition has been generated. Under these conditions, the examination has shown that the Accuver XCAL can easily respond to the $1$-second separated queries, and can track the KPIs in dynamic conditions with high accuracy and a considerable refresh rate. Although the other two methods have relatively moderate to low resolution and refresh rate performances when compared to the Accuver-based method, they might be still preferable as soon as the targeted applications' requirements are satisfied. 

When KPI extraction methods are analyzed comprehensively, the following key insights emerge:
\begin{itemize}
    \item {When ranked from simple to complex in terms of software and hardware requirements, it is observed that the KPI value resolution and data refresh rate improve as the complexity increases. In this regard, the Accuver XCAL-based approach delivers the best performance in terms of these two metrics, whereas the AT-command-based approach can be considered a cost-effective and practical tool within the cost-performance spectrum.}
    \item {For a KPI extraction application as described, it is essential to test the joint usability of the effective method and the network device to be used (CPE, mobile phone, server computer, etc.).}
    \item {It is crucial to select a network device and KPI extraction method that can meet the required KPI value resolution and data refresh rate for the intended application.}
\end{itemize}

The authors believe that the approaches described in this paper will serve as a guide for researchers focusing on innovative, application-oriented studies on 5G infrastructure.

\section*{Acknowledgment}
This study has been supported by the 1515 Frontier Research and Development Laboratories Support Program of TÜBİTAK under Project 5229901 - 6GEN. Lab: 6G and Artificial Intelligence Laboratory. 
The authors express their gratitude to Hakan KOÇAK and Selçuk ATAV from Turkcell, Bünyamin CAN from İris Telecom, and İlkay CİHANER from General Mobile for providing their broadband product GM OD-513 and additional technical supports, to Sabahattin KARAKUŞ, Jaeseon KIM from AANDS Technology Asia Pacific PTE. LTD., R\&D team of MeiG Smart Technology Co. for the technical support with their product MeiG CPE SRT853L, and also to Frank NJIAGA and Abhinav GARG from Accuver Co., Ltd. for their technical support on remote access procedures to their licensed software product XCAL.

\balance

\end{document}